\documentclass[12pt, reqno]{article}

\usepackage{amssymb}
\usepackage{amsmath}
\usepackage{graphics}
\usepackage{float}
\usepackage{graphicx}
\usepackage[margin=1.5in]{geometry}
\usepackage{verbatim}
\usepackage{mathcomp}
\usepackage{caption}
\usepackage{hyperref}

\begin{document}

\begin{center}
{\large \textbf{Repeated noise pattern in the data of arXiv:1807.08572, \\ ``Evidence for Superconductivity at Ambient Temperature and Pressure in Nanostructures''}}

\

\small
Brian Skinner \\
\textit{Department of Physics, Massachusetts Institute of Technology, Cambridge, MA, 02139, USA}
\\
(Dated: \today)
\end{center}

\normalsize 

On July 23, 2018 a preprint appeared on the arXiv reporting the observation of room temperature supercondutivity in a nanostructured solid composed of gold and silver nanocrystals. \cite{paper}  The authors of this preprint present resistance measurements from four samples, and they report an apparent $T_c$ ranging from $145$\,K to greater than $400$\,K.

Given the extraordinary and exciting nature of this claim, it is worth examining the reported data closely.   In this short comment I point out a very surprising feature in the data: an identical pattern of noise for two presumably independent measurements of the magnetic susceptibility.

While the preprint and its supplementary information contain data from four samples, the primary focus in the main text is on a single sample with $T_c = 236$\,K.  In Fig.\ 3(a) of the preprint the authors plot the magnetic susceptibility of this sample as a function of temperature.  Data is plotted for different values of the magnetic field:

\begin{figure}[H]
\centering
\includegraphics[width=0.66\textwidth]{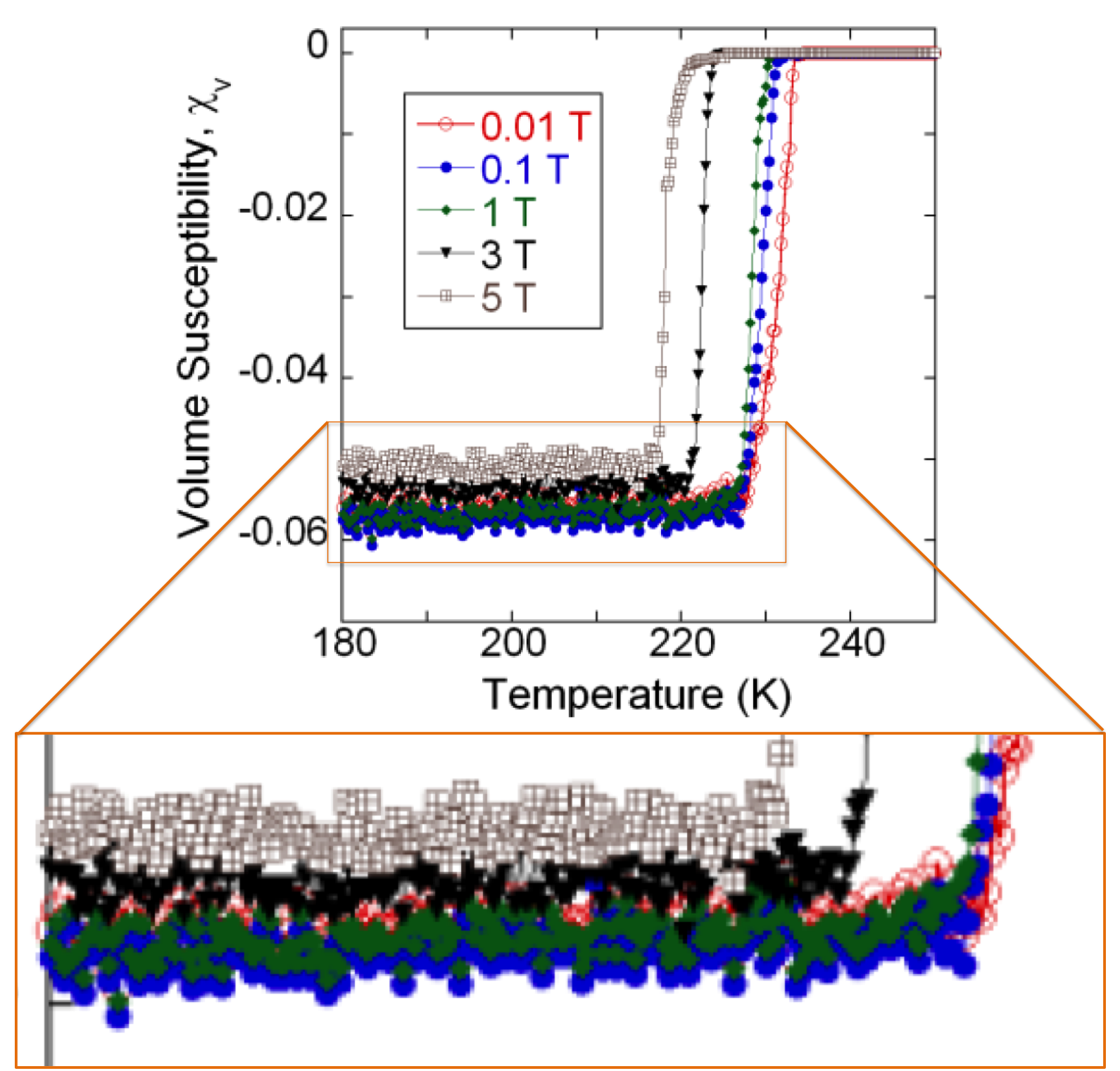}
\captionsetup{labelformat=empty, format=hang}
\begin{quote}
\footnotesize
Figure 3(a) of Ref.\ \cite{paper}, and a zoomed-in view of the data along the lower plateau of susceptibility.  For every green data point at $T \lesssim 225$\,K, there is a blue data point that is displaced downward by a constant amount -- i.e., the two curves have the same pattern of noise.
\end{quote}
\end{figure}

For each curve in this figure, the data points can be divided into two groups.  Data on the lower plateau of susceptibility has relatively large noise while data above this plateau has low noise.  Surprisingly, the data points on the lower plateau appear to be identical for the blue and green curves, with only a constant shift between them.  That is, both curves appear to be described by the same pattern of random noise, with a constant offset separating them.  This correlation in the noise between the two curves disappears abruptly above $T \approx 225$\,K.

It is not easy to tell from the figure whether the red points also have the same pattern of noise.  The white squares and black triangles appear to have a different noise pattern from either the green/blue curves or from each other.

This unusual feature of repeated noise in the magnetic susceptibility has, to my knowledge, no precedent in the superconducting literature, and no obvious theoretical explanation.

\

{\footnotesize
I am grateful to P.~A.~Lee for advising me and encouraging me to post this comment.}



\begin{thebibliography}{100}

\bibitem{paper} D.~K.~Thapa and A.~Pandey, ``Evidence for Superconductivity at Ambient Temperature and Pressure in Nanostructures'', \href{https://arxiv.org/abs/1807.08572}{arXiv:1807.08572} [cond-mat.supr-con] (2018).

\end{thebibliography}
\end{document}